\documentclass[twocolumn,preprintnumbers,amsmath,amssymb]{revtex4}
\usepackage{graphicx}
\usepackage{dcolumn}
\usepackage{bm}
\usepackage{color}

\usepackage{dcolumn}
\usepackage{bm}

\newcommand{\comment}[1]{}

\begin{document}

\title{Photoassociative production of ultracold heteronuclear YbLi* molecules}

\author{Richard Roy}
\author{Rajendra Shrestha}
\author{Alaina Green}
\author{Subhadeep Gupta}
\affiliation{Department of Physics, University of Washington, Seattle, Washington 98195, USA}

\author{Ming Li}
\affiliation{Department of Physics, Temple University, Philadelphia, Pennsylvania 19122, USA}
\author{Alexander Petrov}
\affiliation{Department of Physics, Temple University, Philadelphia, Pennsylvania 19122, USA}
\affiliation{NRC “Kurchatov Institute” PNPI, Gatchina, Leningrad district 188300, Russia;
Division of Quantum Mechanics, St.Petersburg State University, University Embankment 7-9, St. Petersburg, 199034 Russia}
\author{Chi Hong Yuen}
\affiliation{Department of Physics, Temple University, Philadelphia, Pennsylvania 19122, USA}
\affiliation{Alternative address: Department of Physics, University of Central Florida, Orlando, Florida 32816, USA}
\author{Svetlana Kotochigova}
\affiliation{Department of Physics, Temple University, Philadelphia, Pennsylvania 19122, USA}

\date{\today}
\begin{abstract}
We report on the production of ultracold heteronuclear YbLi* molecules in a dual-species magneto-optical trap by photoassociation (PA). The formation of the electronically excited molecules close to dissociation was observed by trap loss spectroscopy. We find 4 rovibrational states within a range of $250\,$GHz below the Yb($^1S_0$) + Li($^2P_{1/2}$) asymptote and observe isotopic PA line shifts in mixtures of $^6$Li with $^{174}$Yb, $^{172}$Yb, and $^{176}$Yb. We also describe our theoretical ab-initio calculation for the relevant electronic potentials and utilize it to analyze and identify the lines in the experimentally observed spectrum.
\end{abstract}
\maketitle

\section{Introduction}
Ultracold heteronuclear molecules have attracted considerable attention because of the prospects for utilizing their rich internal structure as a platform for emulating various many-body Hamiltonians, conducting fundamental studies in quantum chemistry, and performing precision tests of fundamental physics \cite{carr09}. Many of these proposals require the availability of trapped, high phase space density samples of deeply bound ground state molecules. A successful route to such molecules has been demonstrated by first simultaneously cooling two atomic species to high phase space density, subsequently associating free atom pairs into dimers using heteronuclear field-induced Feshbach resonances, and finally de-exciting the resultant molecule using coherent optical multi-photon transitions. To date these methods have been successfully applied to produce trapped samples of four bi-alkali molecules: $^{40}$K$^{87}$Rb \cite{ni08}, $^{87}$Rb$^{133}$Cs \cite{take14,molo14}, $^{23}$Na$^{40}$K \cite{park15}, and $^{23}$Na$^{87}$Rb \cite{guo16}.

While the bialkali systems promise to be a fruitful testbed for exploring dipolar quantum phenomena, it is important to expand the field of ultracold diatomic molecules to different classes of heteronuclear species. Combining alkali and alkaline-earth-like atomic species offers the new property that the $^2\Sigma_{1/2}$ absolute ground state possesses both electric and magnetic permanent dipole moments, which is attractive for applications in quantum simulation and information \cite{mich06}. While cooling to simultaneous quantum degeneracy has been demonstrated in such systems (Yb-Li \cite{hara11,hans11}, Sr-Rb \cite{pasq13}, and Yb-Rb \cite{vaid15}), they inherently suffer from the absence of broad magnetically tunable Feshbach resonances in the ground-state \cite{zuch10}. Such resonances have been verified in an ultracold mixture of ground state alkali Li and excited metastable-state alkaline-earth-like Yb \cite{khra14,dowd15}, but the inelastic losses appear too large to efficiently utilize them for molecule production. A viable alternative to the magnetic Feshbach resonance has recently been demonstrated using coherent optical Raman transfer from free atoms to molecules confined in a three dimensional optical lattice \cite{stel12}. The first step towards realizing this approach in the Yb-Li system is performing spectroscopy of the excited molecular potentials using photoassociation (PA) to determine a suitable intermediate state for the Raman process.

In this work, we demonstrate and characterize interspecies collisions in a dual-species Yb-Li magneto-optical trap (MOT), and carry out PA spectroscopy of molecular potentials below the Yb($^1S_0$) + Li($^2P_{1/2}$) asymptote. We observe 4 rovibrational states in a comprehensive search from 0 to 250 GHz binding energy and measure PA lineshifts for $^6$Li with different isotopes of Yb. We also describe theoretical ab-initio calculations of the relevant ground and excited state potentials. By using these potentials to analyze our experimental observations, we identify the quantum states of the YbLi* molecules produced.

Several challenges were overcome to demonstrate PA spectroscopy in a dual-species Yb-Li MOT. First, it is not possible to utilize a dark-spot MOT \cite{kett93} for either species, a technique that has been very important in earlier MOT fluorescence-based heteronuclear PA experiments \cite{kerm04,wang04,nemi09,dutt14}. For Yb, this results from the lack of hyperfine structure in the electronic ground state. In the case of $^6$Li, unlike other alkali atoms, the small ground state hyperfine splitting renders the lower hyperfine state unusable as a true dark state due to off-resonant excitation from the MOT cooling light. Second, the large transition linewidth mismatch ($\Gamma_{\text{Li}}/\Gamma_{\text{Yb}} \sim 33$) complicates simultaneous MOT operation with a common magnetic field gradient. Finally, the small reduced mass of YbLi results in a large vibrational spacing near dissociation, reducing the number of states with suitably large Franck-Condon factors.

The rest of the paper is organized as follows. Section \ref{sec:experimental-setup} describes the parameters used in our dual-species MOT setup, and characterizes the interspecies interactions in the absence of photoassociation light. In Section \ref{sec:li2PA} we discuss the response of our dual Yb-Li MOT system in the vicinity of Li$_2$ photoassociation resonances. We present our measurements of photoassociative production of YbLi* molecules in Section \ref{sec:liybPA} including isotope shifts of PA lines. In Section V, we discuss the theoretical ab-initio calculations and use it to identify the molecular states observed. We summarize and conclude the paper in Section VI.

\section{\label{sec:experimental-setup} Experimental setup and characterization of dual-species MOT}
Our dual-species Yb-Li experimental apparatus is described in \cite{hans13}. Since neither species can be operated in a dark MOT configuration, we must choose MOT parameters for both Li and Yb that enhance our sensitivity to heteronuclear PA loss, despite the relatively low bright MOT densities. Towards this end, we choose parameters that create a probe and bath situation for Yb and Li, respectively.

\begin{figure}
\begin{centering}
\includegraphics[width=1\columnwidth]{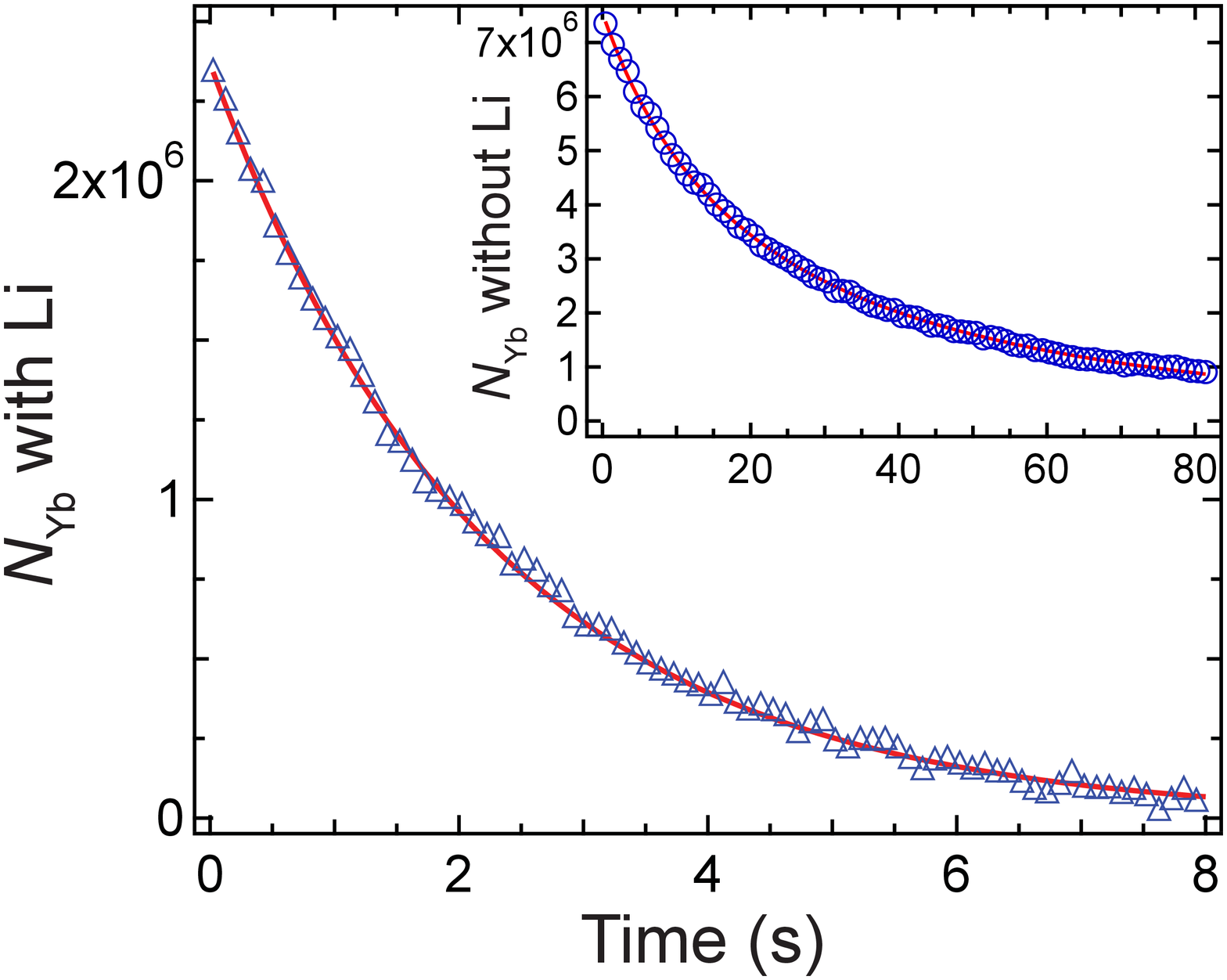}
\par\end{centering}
\caption{\label{fig:fig1} Evolution of Yb MOT number with (open triangles) and without (inset, open circles) Li MOT present (note different time axes). The solid red lines are fits to the data with the solution to equation (\ref{eqn:rateequation1}) with constant Li density, as discussed in the text.}
\end{figure}

Specifically, the shared magnetic field gradient of $B'_{\text{axial}} = 20$ G/cm is chosen to optimize Li number and density. For Li, the cooling (repumping) light parameters for the $F=3/2 \leftrightarrow F'=5/2$ ($F=1/2 \leftrightarrow F'=3/2$) transition are chosen as $\delta_{\text{cool}} = 5\Gamma_{\text{Li}}$ ($\delta_{\text{rep}} = 3\Gamma_{\text{Li}}$) and $I_{\text{cool}}/I_{\text{sat,Li}} = 8$ ($I_{\text{rep}}/I_{\text{sat,Li}} = 0.9$) per beam, where $\Gamma_{\text{Li}}/2\pi = 5.9$ MHz and $I_{\text{sat,Li}} = 2.54$ mW/cm$^2$. To extract the Li cloud density using absorption imaging, we note that the observed in-trap distribution is well fit by a Gaussian. Typical atom numbers and cloud sizes are $N_{\text{Li}} = 7 \times 10^7$, $w_{a,\text{Li}} = 700$ $\mu$m, and $w_{r,\text{Li}} = 900$ $\mu$m, where $w_a$ and $w_r$ are the axial and radial $1/e$ cloud radii, respectively. The corresponding peak Li cloud density is $n_{\text{Li}}(\vec{r}=0) = 2 \times 10^{10}$ cm$^{-3}$. The MOT temperature was measured using time-of-flight absorption imaging and found to be $T_{\text{Li}} = 1$ mK.

Since we operate our Yb MOT on the relatively narrow ($\Gamma_{\text{Yb}}/2\pi = 182$ kHz) $^1S_0$ $\rightarrow$ $^3P_1$ transition at 556 nm, the Yb cloud size at the chosen gradient is considerably smaller than that of Li. Furthermore, the loading rate at a gradient of $B'_{\text{axial}} = 20$ G/cm is much lower than that at our usual operating gradient for Yb of 3 G/cm, resulting in a reduced steady state MOT number. The MOT light parameters used for Yb are $\delta_{\text{Yb}} = 20\Gamma_{\text{Yb}}$ and $I/I_{\text{sat,Yb}} = 160$, where $I_\text{sat,Yb} = 0.14$ mW/cm$^2$, yielding typical numbers and sizes of $N_\text{Yb} = 5 \times 10^6$, $w_{a,\text{Yb}} = 200$ $\mu$m, and $w_{r,\text{Yb}} = 300$ $\mu$m. The resulting peak Yb density is $n_{\text{Yb}}(\vec{r}=0) = 5 \times 10^{10}$ cm$^{-3}$ and temperature $T_{\text{Yb}} = 130$ $\mu$K. The procedure for ensuring optimal overlap of the cloud centers is discussed in Section \ref{sec:li2PA}.

The large mismatch in size and number between the two MOTs allows us to view Yb as a small probe inside of a larger Li bath. In fact, when operating the dual-species MOT, the Li cloud number and size is unchanged, while the steady state Yb number falls by roughly a factor of 3 due to inelastic Li-Yb collisions in the trap. In order to quantify the interspecies collisions, we examine the following rate equations
\begin{align} \label{eqn:rateequation1}
\dot{n}_{\alpha} &= \ell_{\alpha} - \gamma_{\alpha}n_{\alpha} - 2K_{2,\alpha} n_{\alpha}^2 - K_2' n_{\beta} n_{\alpha},
\end{align}
where $\alpha$ = Yb(Li) and $\beta$ = Li(Yb) for the Yb(Li) rate equation, $K_{2,\alpha}$ and $K_2'$ are homonuclear and heteronuclear two-body inelastic loss rate coefficients, respectively, and $\ell_\alpha$ is the spatially dependent loading rate.

The dual-species MOT inelastic dynamics measurements are presented in Figure \ref{fig:fig1}. To extract the interspecies loss rate coefficient, $K_2'$, we map the density evolution from equilibrium of Yb with and without Li present after a quench of the Yb MOT loading rate, $\ell_{\text{Yb}} (t = 0) = 0$, where we block the Yb atomic beam with a mechanical shutter at time $t=0$. We continuously measure the density of Yb by collecting the MOT fluorescence on a photomultiplier tube (PMT) and assuming a constant density profile over the full timescale of decay. This assumption is justified as the Yb cloud widths do not change when the atom number falls by a factor of 3 due to the presence Li. The conversion of PMT voltage to atom number is calibrated using absorption imaging. By integrating equation (\ref{eqn:rateequation1}) for $\dot{n}_{\text{Yb}}$ over space we can relate our measurement of atom number decay $N_{\text{Yb}}(t)$ to density evolution.

We first extract the single species coefficients for Yb in equation (\ref{eqn:rateequation1}) from the decay in the absence of Li, and then fix these parameters in the fit with Li. The red solid line in Figure \ref{fig:fig1} is a fit to the solution of equation (\ref{eqn:rateequation1}), yielding the values $\gamma_{\text{Yb}} = (13.5 \pm 0.3) \times 10^{-3}$ s$^{-1}$ and $K_{2,{\text{Yb}}} = (9.8 \pm 0.3) \times 10^{-13}$ cm$^3$s$^{-1}$, where all quoted errors are statistical. We note that the large initial cloud density and long single body lifetime of $\gamma_{\text{Yb}}^{-1} \approx 74$ seconds allow us to identify both one body and two body loss dominated regimes.

Because of the probe-bath nature of our system, we can replace $n_{\text{Li}}(\vec{r},t)$ in the equation for $\dot{n}_{\text{Yb}}$ by the time-independent peak Li density $n_{\text{Li}}(0)$. For the data in Figure \ref{fig:fig1}, the peak Li density is $n_{\text{Li}}(0) = (1.9 \pm 0.1) \times 10^{10}$ cm$^{-3}$. Performing the same functional fit to the data with Li (open triangles in Figure \ref{fig:fig1}) with the previously determined parameters fixed, we obtain an interspecies loss rate coefficient of $K_2' = (2.3 \pm 0.2) \times 10^{-11}$ cm$^3$s$^{-1}$. Because of the large separation of scale between the interspecies and intraspecies inelastic rates for Yb (see Figure \ref{fig:fig1}) in equation \ref{eqn:rateequation1}, the dynamics in the presence of Li become those of one-body loss with time constant $(K_2' n_\text{Li}(0))^{-1}$, and are thus independent of the Yb cloud shape. Indeed, if we fit the data in Figure \ref{fig:fig1} to the equation $\dot{N}_\text{Yb} = -K_2'n_\text{Li}(0) N_\text{Yb}$, we find the same value for $K_2'$ as quoted above. The above value for $K_2'$ is an order of magnitude larger than that measured in other bright-bright dual-species MOTs \cite{ridi11}. As we will see in Sections \ref{sec:li2PA} and \ref{sec:liybPA}, this has important consequences for our measurements at both Li$_2$ and YbLi PA resonances.

\section{\label{sec:li2PA} L\lowercase{i}$_2$ Photoassociation in a dual Y\lowercase{b}-L\lowercase{i} MOT}
We perform PA spectroscopy on the dual-species MOT by detuning a laser to the red of the Li $^2S_{1/2}$ $\rightarrow$ $^2P_{1/2}$ transition at 671 nm. We are thus sensitive to molecular states in potentials asymptoting to both Yb($^1$S)+Li($^2$P) and Li($^2$S)+Li($^2$P). Since the homonuclear Li($^2$S)+Li($^2$P) interatomic potential falls off as $1/R^3$ at large separation, the background from Li$_2$ PA consists of strong, closely spaced resonances \cite{abra95}. To combat this background, as well as to mitigate the problems of low densities in the bright-bright dual MOT and low Franck-Condon factors due to the $1/R^6$ shorter range potential of Yb and Li, we use a PA beam with very high intensity.

We derive our PA laser beam from a tapered amplifier (TA) system, seeded by a Toptica DL100 external cavity diode laser. The TA output is passed through two identical $670\,$nm band pass filters with a full width at half maximum of 10 nm in order to eliminate frequency components at the $6s6p$ $^3P_1$ $\rightarrow$ $6s7s$ $^3S_1$ transition in Yb at 680 nm. After coupling into a single mode fiber, the PA beam has a power of 100 mW before focusing onto the dual MOT. We leverage the small size of the Yb cloud inside of the Li bath and pass the PA beam through the region of Li-Yb overlap 8 times at an average waist of 80 $\mu$m, with a total peak intensity of 6 kW/cm$^2$.

To optimize overlap of the two clouds, we fix the Yb MOT position and vary the Li MOT position by slight pointing adjustments to the cooling and repumping beams, while monitoring the fluorescence of the Yb MOT. The cloud center can be easily translated across the extent of the Yb cloud, and we find that this has no noticeable effect on Li density. We choose the Li cloud position that maximizes inelastic collisions in the MOT and thus minimizes the Yb fluorescence. This method was checked by monitoring the spatial overlap in fluorescence imaging on 2 independent axes.

For acquisition of PA spectra in the dual-species MOT, we record fluorescence from the Li and Yb MOTs on a photodiode and PMT, respectively, and simultaneously measure the PA laser frequency on a High Finesse WS7/1275 wavemeter. For each scan we sweep the PA laser frequency over a 500 MHz range at 3 mHz, and average over 2 periods.
\begin{figure}
\begin{centering}
\includegraphics[width=1\columnwidth]{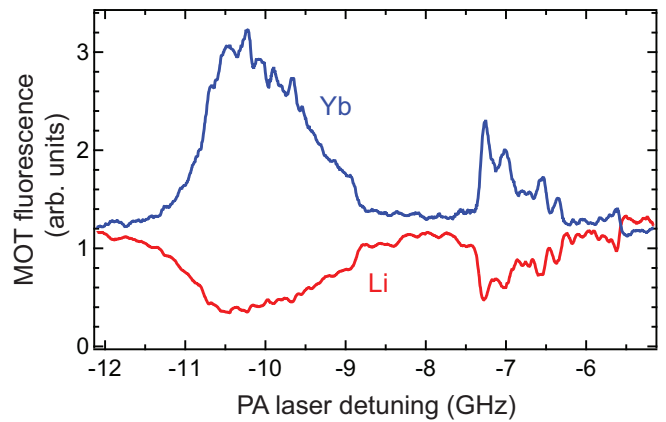}
\par\end{centering}
\caption{\label{fig:lipalines} Dual Li-Yb MOT fluorescence spectra near PA resonances in Li corresponding to the $v = -82$ and $-83$ vibrational states of the $1^3\Sigma_g^+$ potential. Loss of Li (red) due to PA is accompanied by reduction of background inelastic loss for Yb (blue), leading to strong enhancement of the Yb signal.}
\end{figure}

A typical PA spectrum in the vicinity of Li$_2$ PA resonances is shown in Figure \ref{fig:lipalines}, where we reference the laser frequency to the $^1S_{1/2}$ $\rightarrow$ $^2P_{1/2}$ transition. The two broad Li PA features correspond to the $v = -82$ and $-83$ vibrational states of the $1^3\Sigma_g^+$ excited state potential \cite{abra95}. To understand the striking response of the Yb MOT fluorescence around Li PA features, it is useful to examine the spatially integrated rate equation for Yb,
\begin{align}
\label{eqn:rateequation2}
\dot{N}_{\text{Yb}} = L_{\text{Yb}} - \gamma_{\text{Yb}}' N_{\text{Yb}} - \frac{K_2'}{V_{\text{Li}}} N_{\text{Li}}(\delta_{\text{PA}})N_{\text{Yb}}.
\end{align}
Here $L_{\text{Yb}}$ is the loading rate in atoms/sec, $\gamma_{\text{Yb}}'$ is the one-body loss term, which now includes loss due to scattering of PA light on the $6s6p$ $^3P_1$ $\rightarrow$ $6s7s$ $^3S_1$ transition, $V_{\text{Li}} = N_{\text{Li}}/n_{\text{Li}}(0)$, and $N_{\text{Li}}$ is written explicitly as a function of PA laser detuning $\delta_{\text{PA}}$ due to Li$_2$ PA resonances. In equation (\ref{eqn:rateequation2}) we can neglect the term for two-body Yb-Yb collisions because the density in the presence of Li is much reduced, and the one-body loss term now dominates in the presence of PA light.

In the steady state we have $N_{\text{Yb}} = L_{\text{Yb}}/(\gamma_{\text{Yb}}'+K_2' N_{\text{Li}}(\delta_{\text{PA}})/V_{\text{Li}})$, where we find that away from Li$_2$ PA resonances $\frac{K_2' N_{\text{Li}}(\delta_{\text{PA}})}{V_{\text{Li}}\gamma_{\text{Yb}}'} \approx 10$. Hence, the Yb MOT number is inversely proportional to the Li number, and only begins to saturate to its one-body limited value of $L_{\text{Yb}}/\gamma_{\text{Yb}}'$ for large Li loss. As seen in Figure \ref{fig:lipalines}, a unique feature of this probe-bath system is that the Yb signal is greatly enhanced at the positions of Li$_2$ resonances, leading to a detection channel for Li$_2$ PA which increases in signal-to-noise at the location of a resonance.
\section{\label{sec:liybPA} Photoassociative production of Y\lowercase{b}L\lowercase{i}*}
\begin{figure}
\begin{centering}
\includegraphics[width=1\columnwidth]{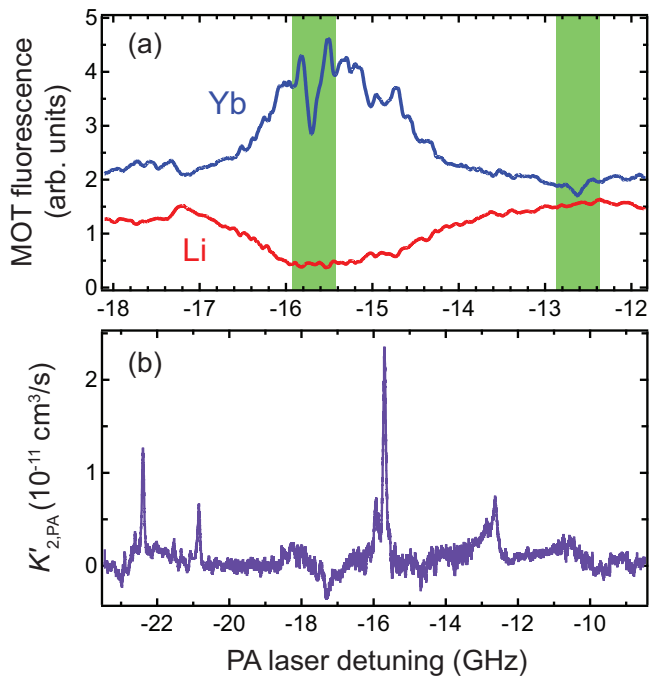}
\par\end{centering}
\caption{\label{fig:fig3} Photoassociation spectra of YbLi*. (a) Dual MOT fluorescence spectra of Li (red) and Yb (blue) revealing 2 YbLi* PA resonances (positions highlighted in green), the stronger of which lies near the center of a broad Li$_2$ PA resonance. (b) Extracted inelastic loss rate coefficient for PA vs. laser detuning, where peaks correspond to YbLi* resonances.}
\end{figure}
We now present our measurements of YbLi* PA resonances in the dual-species MOT. Figure \ref{fig:fig3}(a) shows dual MOT fluorescence spectra, with 2 YbLi* PA resonances visible on top of a broad Li$_2$ PA resonance. Remarkably, even with a 75\% reduction of the Li bath we achieve a contrast of 40\% for the strong resonance at 15.70 GHz detuning.

In order to extract quantitative information about YbLi* resonances independent of the PA frequency-dependent Li number, we express the steady state solution to equation (\ref{eqn:rateequation2}) as
\begin{align}
K_2' &= K_{2,\text{bg}}' + K_{2,\text{PA}}'(\delta_\text{PA}) \nonumber \\
 &= \frac{V_{\text{Li}}L_{\text{Yb}}}{N_{\text{Li}}N_{\text{Yb}}}\left( 1-\frac{\gamma_{\text{Yb}}'}{L_{\text{Yb}}}N_{\text{Yb}}\right),
\label{eqn:betaeqn}
\end{align}
where we have written $K_2'$ as the sum of a constant background inelastic component, $K_{2,\text{bg}}'$, and a PA frequency dependent component, $K_{2,\text{PA}}'(\delta_{\text{PA}})$. Since we can measure the factor $\gamma_{\text{Yb}}'/L_{\text{Yb}}$ from the equilibrium Yb MOT number in the absence of Li, and because we already know the value of $K_{2,\text{bg}}'$ from Section \ref{sec:experimental-setup}, equation (\ref{eqn:betaeqn}) gives us a simple method for producing a spectrum of $K_{2,\text{PA}}'(\delta_{\text{PA}})$ in absolute units (Figure \ref{fig:fig3}(b)).

Each of the 4 YbLi* PA lines shown in the spectrum in Figure \ref{fig:fig3}(b) has a double peak substructure due to the ground state hyperfine splitting in $^6$Li of 228 MHz. The peak YbLi* molecule production rate observed in our experiment is $\sim 5 \times 10^5$ s$^{-1}$, which is comparable to PA experiments in a dark-bright Rb-Yb dual MOT \cite{nemi09}. Any higher-lying (more weakly bound) states are not observed as the fluorescence signal is strongly affected by the proximity to the Li atomic resonances. As these are the only YbLi* structures we see in our scan down to 250 GHz detuning, we can conclude that lower-lying (more deeply bound) vibrational states have Franck-Condon factors that are below our experimental sensitivity.

For the two strongest PA lines at 15.70 GHz and 22.39 GHz, we performed a separate test \cite{nemi09} to verify that our observations correspond to formation of YbLi* and not Yb*Li*. We removed the excited state Yb atoms during PA exposure by pulsing the Yb MOT and PA light 180$^\circ$ out of phase with a $10\,$kHz square wave. Since the Yb excited state lifetime is $\tau = 0.9 \,\mu$s, the observed persistence of the PA features in this situation confirms the production of YbLi*.

\begin{figure}
\begin{centering}
\includegraphics[width=1\columnwidth]{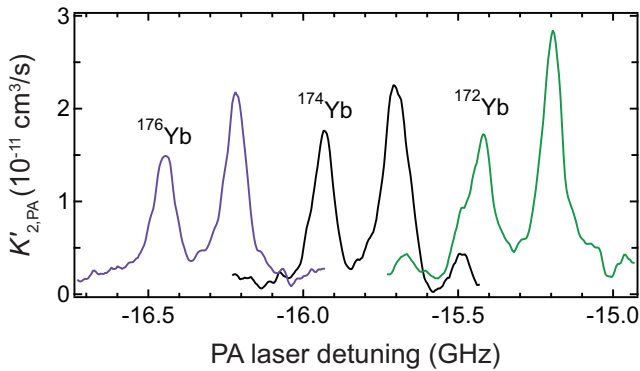}
\par\end{centering}
\caption{\label{fig:isotopedependence} YbLi* photoassociation for different isotopes of Yb. The line shift is due to the changing reduced mass, while the double peak reflects the ground state hyperfine structure in $^6$Li (228 MHz splitting).}
\end{figure}

In order to obtain more information for state assignment to the observed YbLi* lines, we change the Yb isotope in our dual-species MOT and observe the shift in resonance position with changing reduced mass. The measured spectra at the 15.70 GHz PA line for $^{172}$Yb, $^{174}$Yb, and $^{176}$Yb are shown in Figure \ref{fig:isotopedependence}. We find a symmetric shift from the position of the $^{174}$YbLi* resonance of $\pm$510 MHz for the $^{172}$YbLi*$/^{176}$YbLi* resonances, respectively. Similarly, at the 22.39 and 20.89 GHz resonances, we find a shift between the different isotopes of 405 and 390 MHz respectively. The direction of the shift with reduced mass is consistent with an accumulated shift from the most deeply bound to near dissociation molecular states.

\section{Calculation of ab-initio potentials and PA line identification}

The main purpose of our theoretical study is to provide  reliable information about the interaction potentials between Yb and Li atoms in order to identify and analyze the experimentally-observed photoassociation spectrum of the excited YbLi molecule, presented in the previous section. We are interested in all  potentials that dissociate to the limits where both atoms are in their electronic ground state and those where the Li atom is in its first-excited $^2P$ state.

\subsection{Ab-initio calculation of molecular potentials}

Despite the fact that spin-orbit splitting $\Delta$ of the $^2P_{j=1/2,3/2}$ states of Li is  small (here $\Delta/h\approx 10.05\,$GHz \cite{Wall03}, where $h$ is Planck's constant) a precise description of the binding energy of  highly-excited vibrational levels near these dissociation limits (with an accuracy $\delta\epsilon$, where $\delta\epsilon/h$ is a few MHz) requires a relativistic Hund's case (c) description of the long-range excited potentials. There exist two relativistic potentials with $\Omega = 1/2$ symmetry and one with $\Omega=3/2$  that dissociate to the Li $^2P_{j}$ limits. The quantum number $\Omega$ is the  absolute value of the projection of the total electron angular momentum $J$ along the interatomic axis.

We use the {\it ab-initio} relativistic configuration-interaction valence-bond method (RCI-VB) \cite{koto08} to calculate the relevant electronic potentials of Yb$(^1S$)+Li$(^2P$). The ground Yb($^1S_0$)+Li$(^2S_{1/2})$ state potential was calculated in one of our earlier papers \cite{koto11} using a non-relativistic coupled cluster method. We presented its long-range van der Waals coefficient separately in \cite{khra14}. Here, we will slightly modify the short-range form of the ground-state potential to fit to the experimentally measured scattering length \cite{ivan11,hara11} (see the discussion at the end of this Section).

In the valence-bond method electronic wavefunctions $|\Phi;R\rangle$, which are the eigenstates of the relativistic many-electron Hamiltonian $\widehat{H}$ at interatomic separation $R$, are constructed from atomic orbitals of the constituent atoms $at1$ and $at2$. In essence, the electronic wavefunction is given by $|\Phi;R\rangle= \sum_{\alpha\beta}c_{\alpha\beta}|\det_{\alpha\beta}\rangle$, where each $|\det_{\alpha\beta}\rangle$ is an anti-symmetrized  product of  atomic determinants $|\det_{\alpha}^{at1}\rangle$ and $|\det_{\beta}^{at2}\rangle$. Each $|\det_{\gamma}^{at}\rangle$ is  a Slater determinant of atomic Dirac-Fock orbitals. The sums $\alpha$ and $\beta$ are over all included atomic determinants. The variational CI coefficients $c_{\alpha\beta}$ are obtained by solving a generalized matrix eigenvalue  problem as the basis $|\det_{\alpha\beta}\rangle$ is not orthogonal. We use  atomic Dirac-Fock orbitals that are optimized iteratively to  accelerate the CI convergence. This includes alternating between the determination of CI coefficients and the improvement of the atomic orbitals.

Our RCI-VB method is an  {\it all-electron} calculation in which no pseudopotentials are used. Instead the orbitals (or electrons) of each atom are divided into two groups: core and valence electrons. Here core electrons occupy the 1$s^2_{1/2},\dots$, 4$d_{5/2}^{10}$ closed shells of the Yb atom, where the superscript denotes the number of electrons in orbital $n \ell_j$.  Core orbitals  do not contribute significantly to the formation of the molecule and excitations out of these shells are not considered. Valence electrons occupy the 5$s^2_{1/2}$, 5$p^2_{1/2}$, 5$p^4_{3/2}$, 4$f^{6}_{5/2}$, 4$f^{8}_{7/2}$, and 6$s^2_{1/2}$ orbitals for the Yb atom and the 1$s^2_{1/2}$ and 2$s^1_{1/2}$ orbitals for Li. 
The convergence of the CI expansion can only be determined by increasing the number of configurations. We achieved this by systematically adding atomic orbitals out of which the atomic and thus molecular determinants are constructed. In particular, we allow single and double excitations from valence shells into virtual, unoccupied orbitals with principal quantum numbers up to $n=5$ for Li and $n=8$ for Yb, and orbital quantum numbers $\ell=0, 1, 2,$ and $3$.

\begin{figure*}
\includegraphics[scale=0.3,trim=0 33 0 40,clip]{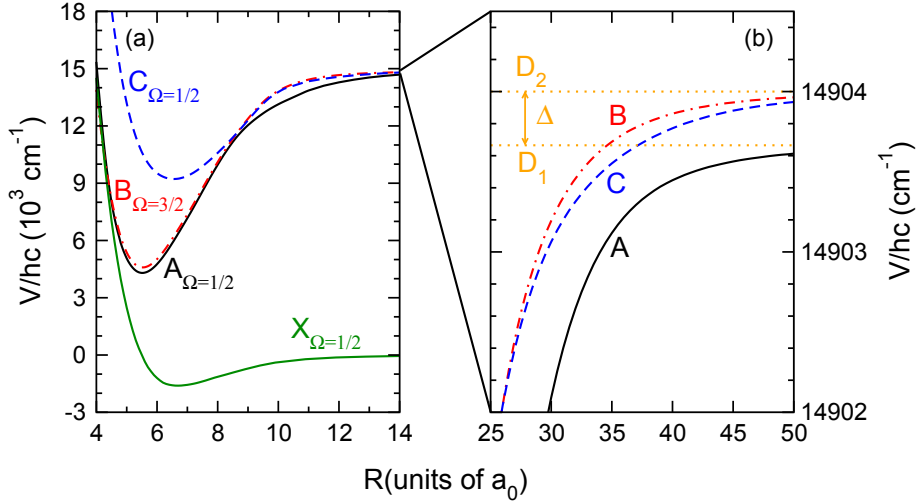}
\caption{(a): Relativistic ground X($\Omega$=1/2) (dark green solid line) and excited adiabatic A($\Omega$=1/2) (solid black line), B($\Omega$=3/2) (red dash-dotted line), and C($\Omega$=1/2) (blue dashed line) potentials of Yb($^1S_0$)+Li($^2P_{1/2,3/2}$) as a function of the internuclear separation $R$.  (b): The three excited potentials  that dissociate to the D$_1$ ($^2P_{1/2}$) and D$_2$ ($^2P_{3/2}$) limits of Li at large $R$.
}
\label{pots}
\end{figure*}

Our  observed photoassociation spectrum is localized within 250 GHz of the $^2P_{1/2}$ or D$_1$ dissociation limit of Li. Consequently, we need to pay  particular attention to the long-range molecular dispersion potentials.  We have  performed a separate calculation of the  dispersion interaction potential for this excited YbLi molecule and have connected it smoothly to the electronic calculations discussed above. We use  degenerate second-order perturbation theory
in the anisotropic electric dipole-dipole interaction similar to that described by \cite{ston96}.

We start by defining basis states for the two atoms.  For Li we can first ignore the weak spin-orbit interactions and define non-relativistic states $|n_1 l_1 \Lambda_1 s_1 \Sigma_1\rangle$, where $\vec l_1$ and $\vec s_1$ are its  electron orbital angular momentum and spin, respectively. The corresponding projections $\Lambda_1$ and $\Sigma_1$ are along the internuclear axis and index $n_1$ further labels the atomic states. The energy of these states, $E_{n_1l_1}$, corresponds to the so-called barycenter of the $^{2s+1}l_j$ Li limits and only depends on $n_1$ and $l_1$. Similarly, non-relativistic electric-dipole matrix element $\langle n_1 l_1 \Lambda_1  | d_1 | n_a l_a \Lambda_a \rangle$ are constructed from observed or computed dipole moments assuming  L-S or Russell-Sanders coupling  of the electron orbital angular momentum and spin. They are independent of the electron spin. On the other hand, for Yb,  relativistic corrections are important and we define atomic eigenstates $|n_2 j_2\Omega_2\rangle$ with  energies $E_{n_2j_2}$, where $\vec\jmath_2$ is the total electronic angular momentum of Yb with projection quantum number $\Omega_2$ along the internuclear axis. The index $n_2$ further labels its atomic states.

Our hybrid  zeroth-order basis functions $|\Lambda_1,\Sigma_1,\Omega_2\rangle\equiv |n_1 l_1 \Lambda_1 s_1 \Sigma_1\rangle |n_2 j_2\Omega_2\rangle$ have $l_1=1$  and $s_1=1/2$ for $n_1={\rm Li}(^2$P) and $j_2 = 0$ (and thus $\Omega_2=0$) for $n_2={\rm Yb}(^1$S). We denote their energies by $E_1$ and $E_2$, respectively.
Second-order perturbation theory in the dipole-dipole interaction then leads to the long-range potential  $-\hat C_6/R^6$, where $\hat C_6$ is an operator with matrix elements
\begin{equation}
\langle \Lambda_1,\Sigma_1,\Omega_2|\hat C_6| \Lambda'_1,\Sigma'_1,\Omega'_2\rangle =
 \delta_{\Sigma_1,\Sigma'_1} \sum_{l_a,j_b} K^{l_1j_2}_{l_aj_b} A^{l_1j_2l_aj_b}_{\Lambda_1\Omega_2,\Lambda'_1\Omega'_2}\,,
\end{equation}
where $\delta_{i,j}$ is the Kronecker delta,
\begin{eqnarray}
	K_{l_aj_b}^{l_1j_2} =\frac{1}{(4\pi\epsilon_0)^2}  \sum_{n_a,n_b}'
	\frac{\left|\left<n_1l_1||d_1||n_al_a\right>\left<n_2j_2||d_2||n_bj_b\right>\right|^2}
	{(E_{n_al_a}+E_{n_bj_b}) - (E_1+E_2)}\,,\nonumber
\end{eqnarray}
functions $\langle \cdot || d_i || \cdot \rangle$ are reduced matrix elements of  atomic electric dipole operators, the sums are over all pairs $n_a,n_b$ excluding the zeroth-order state $n_1,n_2$, and $\epsilon_0$ is the electric constant. The angular function
\begin{eqnarray}
\lefteqn{A^{l_1j_2l_aj_b}_{\Lambda_1\Omega_2,\Lambda'_1\Omega'_2}  =  \sum_{\Lambda_a,\Omega_b} (1+\delta_{\Lambda_1,\Lambda_a} )(1+\delta_{\Lambda'_1,\Lambda_a} ) } \\
& &
\times \left(\begin{array}{ccc}
                        l_1 &    1      & l_a \\
                        -\Lambda_1 & (\Lambda_1-\Lambda_a) & \Lambda_a
      \end{array}\right)
\left(\begin{array}{ccc}
                          j_2 &    1      & j_b \\
                         -\Omega_2 & (\Omega_2-\Omega_b) & \Omega_b
       \end{array}\right)
       \nonumber \\
&&
\times\left(\begin{array}{ccc}
                           l_a &    1      & l_1 \\
                          -\Lambda_a & (\Lambda_a-\Lambda'_1) & \Lambda'_1
       \end{array}\right)
\left(\begin{array}{ccc}
                           j_b &    1      & j_2 \\
                          -\Omega_b & (\Omega_b-\Omega'_2) & \Omega'_2
        \end{array}\right)  \nonumber
\end{eqnarray}
is independent on atomic transition frequencies and dipole moments and  $({\cdots\atop\cdots})$ are Wigner-3j symbols. The $\hat C_6$ operator conserves $\Omega=\Lambda_1+\Sigma_1+\Omega_2$ as expected from the selection rules of the dipole-dipole interaction.

We have determined the two independent $K_{l_aj_b}^{l_1j_2} $ using 13 experimental transition frequencies and oscillator strengths from the D$_1$ and D$_2$ excited states of the Li atom \cite{wies09}. We also calculated electronic dipole moments and frequencies to a few highly excited S and D levels using the configuration interaction method of Ref. \cite{koto07} in order to accurately reproduce the dynamic scalar and tensor polarizability of Li($^2P$) computed in Ref.~\cite{tang10}.  For Yb we have used 22 experimental transition frequencies and oscillator strengths starting from the $^1S$ ground state \cite{koma91,blag94,bowe99,dzub10,belo12}. The two independent $K_{l_aj_b}^{l_1j_2}$ are $K_{0~1}^{1~0} = 9602\,E_ha_0^6$, $K_{2~1}^{1~0} = 6586\,E_ha_0^6$, where $E_h$ is the Hartree energy and $a_0=0.0529$ nm is the Bohr radius.

By inspection we further see that in the hybrid product basis  the $\hat C_6$ operator  is  diagonal in all  quantum numbers and, in fact, its matrix elements only depend on the absolute value of the electronic orbital angular momentum, i.e. $|\Lambda_1|=0,1$ or more conveniently $\Sigma$ and $\Pi$. After evaluating the angular function we find  $C_6(\Sigma) = 5877\,E_ha_0^6$ and $C_6(\Pi) = 2457\,E_ha_0^6$.

Next we  include the spin-orbit interaction of Li($^2P$) in the description of the long-range interactions by following Ref.\cite{movr77}. We define the Hund's case (c) basis set
\begin{equation}
|j\Omega\rangle = \sum_{\Lambda_1\Sigma_1} |\Lambda_1\Sigma_1, \Omega_2= 0\rangle \langle l_1 \Lambda_1 s_1\Sigma_1| j\Omega\rangle\,, \label{trans}
\end{equation}
where $\vec\jmath=\vec l_1+\vec s_1+\vec\jmath_2$ is the total molecular electronic angular momentum with a projection $\Omega$ on the internuclear axis, we used that the atomic angular momentum of Yb is zero ($j_2=\Omega_2=0$), and $\langle j_1 m_1 j_2 m_2|j_3m_3\rangle$ is a Clebsch-Gordan coefficient. There are three channels in this basis. The channel $|j\Omega\rangle=|3/2, 3/2\rangle$ does not interact with the other two and its long-range potential is
\begin{equation}
V_{\Omega=3/2}(R) = \Delta  - \mathrm{C}_6(\Pi)/R^6\,.
\end{equation}
The two $\Omega = 1/2$ channels, $|j\Omega\rangle=|1/2, 1/2\rangle$ and $|3/2, 1/2\rangle$, do couple. Their long-range interaction is found by adding the spin-orbit interaction, which is diagonal in the Hund's case (c) basis, and the dispersion potential, which is diagonal in the hybrid basis and must be  transformed with Eq.~\ref{trans} into the $|j\Omega\rangle$ basis. The final long-range potential matrix is
\begin{widetext}
\begin{equation}
V_{\Omega = 1/2}(R) =
\left(
\begin{array}{cc}
0 & 0 \\
0& \Delta
\end{array}
\right)
 +\frac{1}{3} \frac{1}{R^6}
\left(
\begin{array}{cc}
- [C_6(\Sigma)+2C_6(\Pi)] & \sqrt{2}[C_6(\Sigma)-C_6(\Pi)] \\
\sqrt{2}[C_6(\Sigma)-C_6(\Pi)] &-[2C_6(\Sigma)+C_6(\Pi)]
\end{array}
\right) \;.
\end{equation}
\end{widetext}

Diagonalization of this potential matrix leads to  two adiabatic potential curves for $\Omega = 1/2$. The adiabatic $\Omega=3/2$  and the two $\Omega = 1/2$ potentials have been smoothly connected to our {\it ab initio} potentials at $R=16.5\,a_0$ by adjusting, within 5\%, the depth of the potential wells as well as adding  adjustable long range $-C_8/R^8$ potentials in the hybrid basis.

\begin{table}[b]
\caption{Molecular spectroscopic constants for $^{174}$Yb$^6$Li* of relativistic electronic
potentials dissociating to the Yb($^1S$)+Li($^2P$) limits. Here, $D_e$ is the dissociation energy, $\omega_e$
is the  harmonic frequency, and $R_e$ is the bond length. The  notation for the states is defined in Fig.~\ref{pots}.}
\begin{tabular}{cccc}
\hline
State &   $D_e/(hc)$    & $\omega_e/(hc)$ & $R_e$   \\
         & (cm$^{-1}$)& (cm$^{-1}$)& $(a_0$) \\
\hline
A ($\Omega=1/2$)   & 10763       &  317          &   5.52      \\
B ($\Omega=3/2$)  &  10404       & 313           &   5.53         \\
C ($\Omega=1/2$)  &  5714        & 192           &   6.55        \\
\hline
\end{tabular}
\label{constants} \end{table}

The resulting three  excited potentials  dissociating to the Yb($^1S_0$)+Li($^2P_{1/2,3/2}$) limits as well as the ground-state potential are presented in Fig.~\ref{pots}. The excited $\Omega$ = 1/2  potentials, one dissociating to the Yb($^1S_0$)+Li($^2P_{1/2}$)  and one to the Yb($^1S_0$)+Li($^2P_{3/2}$)  atomic limit, have  one avoided crossing located at $ R = 8.5\,a_0$. It is unique to our heteronuclear system. For $R > 40\,a_0$ the three adiabatic potentials approach their dissociation limit with a dispersion potential but now with coefficients  $C_6= 2457\,E_h a_0^6$, $ 4737\,E_h a_0^6$, and $3597\,E_h a_0^6$ for the $\Omega=3/2$ potential and the $\Omega=1/2$ potentials going to the Li($^2P_{3/2}$) and Li($^2P_{1/2}$) limit, respectively. The spectroscopic constants  for the excited potentials are given in Table \ref{constants}. These constants have also been calculated in Ref.\cite{gopa10} using a pseudo-relativistic approach at the spin-orbit CASPT2 level. We agree with their values of $D_e$, $\omega_e$ and $R_e$ within 2\%, 10\%, and 1\%, respectively. The ground state X potential has an
attractive long-range dispersion potential with coefficient $C_6({\rm X}) = 1496(150)\,E_ha_0^6$.

\begin{table}
\caption{\label{tb:exp_data} Position and identification of the observed Yb$^6$Li* photoassociation resonances (see Figure~\ref{fig:fig3}(b)). The first column indicates whether a row shows (E)xperimental or (T)heoretical data. The second column shows the binding energy $\Delta_{174}$ of the PA lines for the $^{174}$Yb $^6$Li* isotope relative to  $D_1$ limit. Column three shows the isotope shift $\delta_{176-174}=\Delta_{176}-\Delta_{174}$ for  $ ^{176}$Yb$^6$Li*. The experimental uncertainty is $10\,$MHz, stemming from the accuracy of the wavemeter. The last three columns present our identification of each resonance. These are the electronic potential A, B or C as defined in Fig.~\ref{pots}, the total angular momentum $J$, and vibrational quantum number, $v$, counted  down from the dissociation limit, respectively.}

\begin{ruledtabular}
\begin{tabular}{llccccc}
  &$\Delta_{174}/2\pi$ & $\delta_{176-174}/2\pi$ & Potl. & $J$ & $v$ \\
  & (GHz) &    (MHz)  &   & \\
  \hline
E &$-12.63$ & $-$ &   &  &  \\
T     &$-12.63$& $ -512$   & B & 3/2 & -2 \\ \hline
E &$-15.70$ & $-510$ &   &     &    \\
T     &$-15.71$& $-513$   & A & 1/2 & -2 \\ \hline
E &$-20.84$ & $-390$ &   &     &    \\
T     &$-20.94$& $-398$   & C & 3/2 & -2 \\  \hline
E &$-22.39$ & $-405$ &  &     &    \\
T     &$-22.26$& $-405$  & C & 1/2 &-2
\end{tabular}
\end{ruledtabular}
\end{table}

A notable feature of the adiabatic potentials is the avoided crossing between the two $\Omega=1/2$ potentials at  $R\approx 8.5\,a_0$. At this point the two potentials are separated by $\delta E/(hc)=230\,$cm$^{-1}$. A Landau-Zener analysis~\cite{land77} assuming a total energy close to the D$_1$ dissociation limit shows that the crossing is close to diabatic. The Landau-Zener transition probability between the two adiabatic surfaces at the crossing point is higher than $99\%$. Thus, for the analysis of our spectroscopic data we modified or diabatized the short range $\Omega=1/2$ potentials to cross each other. We swapped the potentials on the outside of the crossing to calculate the isotope shifts of the observed resonances, while maintaining the labeling of the potentials on the inside of the crossing in Fig.~\ref{pots}(b).

\subsection{Comparison to observation and PA line identification}

The temperature of both atomic ensembles in our experiment is $\leq 1\,$mK and, consequently, the initial collision on the X potential is dominated by
a single value of the relative orbital angular momentum $\vec \ell$, the $\ell=0$ or $s$ partial wave. This implies that the total molecular angular momentum
$\vec J=\vec \ell +\vec \jmath$ equals 1/2 as for the collision $l_1=0$, $s_1=1/2$, and $j_2=0$. Selection rules for a electric dipole transition then ensure that the total angular momentum of the excited molecular state must either be $J = 1/2$ or $3/2$. (This analysis ignores the effects of the nuclear spin of the $^6$Li atom. The approximation is valid as the hyperfine splittings of the excited Li atom are small and not resolved in our experiment.)

The observed PA resonances and isotope shifts are shown in Table~\ref{tb:exp_data}.  Our assignment of these ro-vibrational states is primarily based on matching the observed isotope shifts with those computed for vibrational levels of the three excited diabatized or crossing potentials, but also on the well-characterized spacing of vibrational levels and rotational constants based on our accurate knowledge of dispersion coefficients. Finally, we rely on an analysis of the dependence of vibrationally-averaged dipole moments (VADM) between the initial scattering wavefunction and excited ro-vibrational states on vibrational level, $v$. The dependence of the VADM on $v$ can explain the absence of the observation of more deeply-bound PA lines.

Then we slightly adjusted the depth of the potential wells as well as adding additional long range $-C_8/R^8$ terms to match the exact energy levels and isotope shifts.

The spacing between vibrational states, predominantly determined by the accurate dispersion coefficient, are known sufficiently well to confirm that the observed vibrational state is $v=-2$. Using the adjusted potentials, we calculated the vibrationally averaged dipole moments (VADM) between the ground continuum state and the four observed excited rovibrational states.

For the ground state potential, we made sure that the short-range potential reproduces the $s$-wave scattering length within the range given by previous experiments ($13\pm 3~a_0$) and ($18.9 \pm 3.8~ a_0$) \cite{ivan11,hara11,footasign}. The asymptotic energy of the ground continuum state is taken to be $1$ mK. The $R$-dependent electronic transition dipole moments are from \cite{gopa10}. The final assignment of the PA resonances is shown in Table \ref{tb:exp_data}.

\begin{figure}
\includegraphics[width=\linewidth]{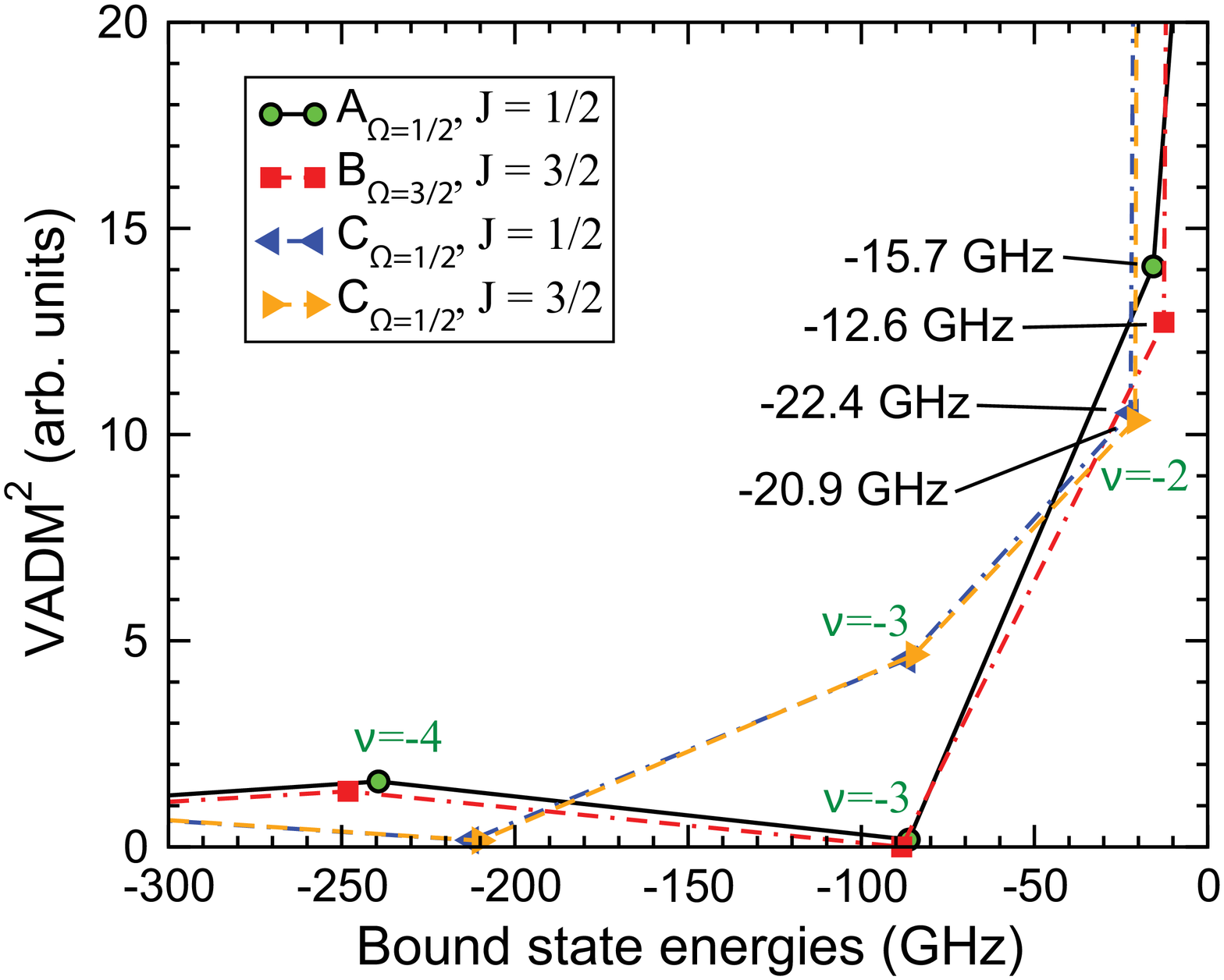}
\caption{Square of vibrationally averaged dipole moment (VADM$^2$) as a function of bound state energy for the vibrational levels of the excited A, B, and C potentials. The scattering length and $C_6$of the ground state potential are $+16\,a_0$ and $1646\,E_ha_0^6$, chosen from within earlier experimental and theoretical bounds, to best reproduce the relative strength of the observed resonances in Fig. 3b. The dipole moments for the $v = -1$ states are orders of magnitude larger than the ones shown on the graph.}
\label{fig:dip_250}
\end{figure}

The range of $-250$ GHz below the $^6$Li D$_1$ line must include vibrational levels with the quantum numbers from $v = -1$ to $v = -4$. The $v = -1$ states were not detected as they are located in the area where the fluorescence signal is dominated by the Li atomic resonances. Since the four $v = -2$ resonances are the only ones observed within this region, we expect that the VADMs for the $v = -3$ and $-4$ bound states should be significantly smaller than the ones for the $v = -2$ bound states. This is the case for the scattering length larger than $16\,a_0 $, which the upper limit of the value, measured in \cite{ivan11} and lower limit of \cite{hara11}.

Figure~\ref{fig:dip_250} illustrates the calculated VADMs for the rovibrational states going as deep as $-300$ GHz from the D$_1$ line, with the ground state scattering length chosen to be $16\,a_0$. It is evident that the more deeply lying states have significantly smaller VADMs comparing to the $v = -2$ ones, which supports the experimental observation.

\section{\label{sec:conlusion} Conclusions}
We have developed and characterized a bright-bright Yb-Li dual MOT suitable for photoassociation studies in the Yb probe, Li bath regime. We observe both one-body and two-body loss dominated regimes in the loss dynamics of Yb in the absence of Li. Our determination of background interspecies inelastics allows us to extract PA loss rate coefficients from equilibrium fluorescence measurements alone. We locate 4 YbLi* PA resonances and measure line shifts from changing the Yb isotope. In order to analyze the molecular states observed, we have performed a theoretical ab-initio calculation for the relevant ground and excited potentials. We have used these potentials to understand our experimental observations and identified the quantum states of the YbLi* molecules produced.

These studies form the starting point for realizing an all-optical coherent production method of ground state YbLi molecules by 2-photon photoassociation. Future studies include spectroscopy of the ground state interatomic potential by 2-photon PA spectroscopy. Ultimately, strong confinement in a three dimensional optical lattice with unit filling of Yb and Li should allow efficient production of ground state molecules.

\section{Acknowledgments} Research at University of Washington was supported by NSF Grant No. PHY-1306647, AFOSR Grant No. FA9550-15-1-0220 and ARO MURI Grant No. W911NF-12-1-0476. Research at Temple University was supported by ARO MURI Grant No. W911NF-12-1-0476, AFOSR Grant No. FA9550-14-1-0321 and NSF Grants No. PHY-1619788 and No. PHY-1125915.


\begin{thebibliography}{41}
\bibitem[{foo(2016)}]{footasign}
\bibinfo{journal}{We have verified the positive sign of the YbLi ground state
  scattering length in an experiment which measures the interspecies mean field
  interaction through collective oscillations in the quantum degenerate regime
  \cite{roy16}}.
\end{thebibliography}

\end{document}